\documentclass[aps,twocolumn,epsfig,showpacs]{revtex4}
\usepackage{graphicx}
\usepackage{epsfig}
\newcommand{\be}{\begin{equation}}
\newcommand{\ee}{\end{equation}}
\newcommand{\bea}{\begin{eqnarray}}
\newcommand{\eea}{\end{eqnarray}}

\newcommand{\eu}{\epsilon_u}

\begin{document}
\title{Role of pulling direction in understanding the energy landscape 
of proteins}
\author{R. Rajesh}
\affiliation{Institute of Mathematical Sciences, CIT Campus, Taramani, 
Chennai-600113, India}
%\ead{rrajesh@imsc.res.in}
\author{D. Giri}
\affiliation{Department of Applied Physics, IT, Banaras Hindu University,
Varanasi 221 005, India}
%\ead{dgiri@bhu.ac.in}
\author{I. Jensen}
\affiliation{Department of Mathematics and Statistics,
The University of Melbourne, Victoria 3010, Australia}
%\ead{I.Jensen@ms.unimelb.edu.au}
\author{S. Kumar}
\affiliation{Department of Physics, Banaras Hindu University,
Varanasi 221 005, India}
%\ead{yashankit@yahoo.com}
\date{\today}

\begin{abstract}
Single molecule force spectroscopy provide details of the underlying 
energy surfaces of proteins which are essential to the understanding 
of their unfolding process. Recently, it has been observed experimentally 
that by pulling proteins in different directions relative to their 
secondary structure, one can gain a better understanding of the shape 
of the energy landscape. 
We consider simple lattice models which are anisotropic in nature 
to study the response of a force in unfolding of a polymer. 
Our analytical solution of the model, supported by extensive numerical 
calculations, reveal that the force temperature diagrams are very different 
depending on the direction of the applied force. We find that either unzipping 
or shearing kind transitions dominate the dynamics of the unfolding process
depending solely on the direction of the applied force. 
\end{abstract}
\pacs{64.90.+b,36.20.Ey,82.35.Jk,87.14.Gg }
\maketitle

\section{Introduction}

Recent experiments have shown that the energy landscapes governing the
mechanical unfolding of proteins are different from the energy landscapes 
governing the thermal or chemical unfolding 
\cite{rief,busta,tskhovrebova,busta1,carrion1}. 
Until recently it was believed that all residues of any given protein
contribute equally to the thermodynamic stability of globular
proteins, with the stability of the protein structure 
being provided by its hydrophobic core \cite{serrano,eriksson}. 
It was therefore 
expected that the  mechanical unfolding of a protein should be insensitive 
to the direction of the applied force. Surprisingly, recent experiments have 
shown that a protein's resistance to unfolding depends strongly on 
the pulling direction \cite{brockwell,carrion,li,matouschek}. Varying the
pulling direction also gives Angstrom precise information about the 
structure of biomolecules \cite{dietz,dietz1}.

Force extension curves obtained by single molecule experiments (unfolding of proteins, 
stretching of DNA) are often described by the worm like chain model (WLC model) 
or freely jointed chain model (FJC model) \cite{fixman,doi}. The mechanical properties of the FJC 
and WLC models are  well understood, and the usefulness of these models
stems from their simplicity, which means that the analytic expression for the extension
can be written in a very simple form  \cite{Marko,rief98}. 
The only parameter appearing in the description of these models is the persistence length 
of the polymer chain. The force-extension curves obtained from these models 
\cite{Marko, rief98,degennes} show
an excellent agreement with the main features observed in 
early experiments \cite{rief,busta,tskhovrebova,busta1}.
Since the FJC and WLC models are isotropic, in order to study the resistance of a protein 
to unfolding when the pulling geometry is changed, one would have to use different persistence 
lengths (along different directions) to fit the force extension curves \cite{brockwell,carrion,rief98}. 
So when a polymer has been pulled along the $x$-direction  one may use a certain persistence
length which reproduces the observed force-extension curve.
If the pulling direction is changed another persistence length is used to fit
the new force-extension curve. However, the underlying model is still isotropic and
the procedure is somewhat ad hoc and does not provide a unified approach to the
modeling of the anisotropic behavior \cite{creighton,rosa} seen in experiments. Equally importantly
these simple models do not incorporate excluded volume effects in its description.

In a recent paper Kumar and Giri \cite{kumar} used partially directed
self-avoiding walks (PDSAWs) to model anisotropic  biomolecules. 
When a force is applied to one end of the chain it undergoes a transition 
from a folded to an unfolded state. 
Numerical studies  based on a chain size of $N=30$ showed that when the chain 
is being pulled along the preferred direction of the polymer, 
the nature of the transition is akin to unzipping, 
but when force is applied perpendicular to the preferred direction,
the transition is akin to shearing. 

In section $II$  we provide the
analytical solution of the model and show that a change in the
pulling direction gives rise to different phase diagrams even in 
the thermodynamic (infinite chain length) limit. In section~ $III$
we report on an algorithmic breakthrough which has enabled us to increase 
the chain length to $N=150$ which is five times longer
than in the previous study \cite{kumar}. The longer
series are used to obtain numerical estimates for the phase boundary
in excellent agreement with the exact results. In this section we also discuss 
the relevance of two ensembles namely, the constant force ensemble (CFE) and the constant 
distance ensemble (CDE) which are appropriate when describing different experimental setups. 
With the enhanced information about the exact density of states of longer chains, 
we are able to study  precisely finite-size effects which are crucial to 
all single molecule experiments. In particular we find marked differences
between the cases where the force is applied along the $x$- or $y$-directions,
respectively, and we explain these observed differences using simple heuristic
arguments.  The PDSAW is a very restricted model and in order to
further examine the role of anisotropy we introduce an anisotropic
self-attracting self-avoiding walk (ASASAW) model and using  exhaustive
exact enumeration data for chains up to length $N=48$ study the combined 
effect of anisotropy and pulling in section $IV$.
The paper ends with a brief discussion in section $V$.

\section{PDSAW Model and Phase Diagram}

\begin{figure}
\begin{center}
\includegraphics[width=6.0cm]{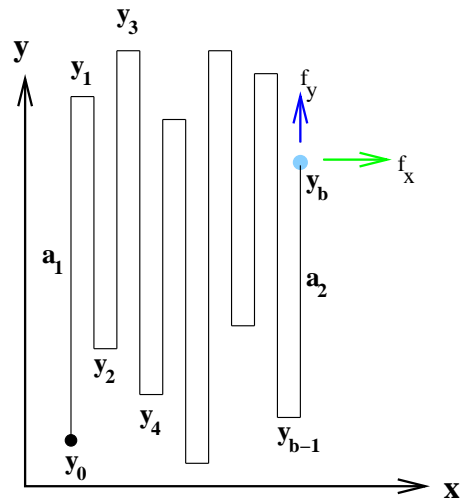} 
\end{center}
\caption{\label{fig:pdsaw} (Color online) Schematic representation
of a PDSAW on the square lattice with one end kept fixed
and the other end subjected to a pulling force along ($f_x$) 
and  perpendicular ($f_y$) to the preferred direction. At low temperature 
the conformation mimics a $\beta$-sheet.
}
\end{figure}

The PDSAW is a self-avoiding walk in which steps along some direction 
(say, the negative $x$-direction) are forbidden. We study this model on a two
dimensional square lattice. An attractive energy $\eu$ is associated with
each non-bonded nearest neighbor \cite{vander}.  At very 
low temperatures the attractive interactions dominate,
and the PDSAW is in a collapsed phase, 
where the density of monomers in the bulk 
is close to one. The typical configurations of the chain mimics the 
$\beta$-sheet \cite{creighton} (see Fig.~\ref{fig:pdsaw}).
At high temperatures thermal fluctuations are strong enough
to break some of the interaction bonds and with increasing entropic 
contributions to the free energy the chain can be in an extended
phase.
A force $\vec{F}$ may be applied along some fixed direction 
giving rise to a stretching energy $E_s$ arising from the
applied force $\vec{F}$ 
\begin{equation}
E_s = -\vec{F}\cdot \vec{\alpha}
\end {equation}
where $\vec{\alpha}$ is the end-to-end vector. In this paper we
consider only the cases where the force is applied either along
the $x$-direction ($f_x$) or the $y$-direction ($f_y$).

The phase boundary between the collapsed and the extended 
phase can be obtained by calculating the macroscopic shape of the collapsed 
phase at low temperatures. 
Let $\omega_x = \exp(f_x/T)$,
$\omega_y = \exp(f_y/T) $ and
$u = \exp(\eu/T)$, where $T$ is the temperature. We set the Boltzmann constant
$k_{B}$ equal to $1$.
The energy of the configuration which mimics
the $\beta$-sheet is (see Fig.~\ref{fig:pdsaw})
\be
 E = -\eu N + \frac{\eu}{2}(a_1 + a_2+ 2 b) +\frac{\eu}{2}
\sum_{j=0}^{b-2} | y_{j+2}-y_{j} |,  \label{eq:12}
\ee
where, as illustrated in  Fig.~\ref{fig:pdsaw},  $a_1=|y_1-y_0|$ is the height (number of steps 
in the $y$-direction) of the first column, $a_2=|y_b-y_{b-1}|$ is the height of the last column 
and $b$ is the number of steps in the $x$-direction.

Following the approach developed in \cite{suppl}, we express the sum
appearing in Eq.~(\ref{eq:12}) as a surface tension term. The macroscopic
shape of the polymer is calculated and the value of $\omega_x$ at which
the collapsed phase becomes unstable gives the critical value,
$\omega_x^c$, in terms of $\omega_y$ and $u$:
\be \label{eq:pbg}
\omega_x^c = \frac{u}{(u-1) \sqrt{\omega_y}} \left[ (\sqrt{u}-
\sqrt{\omega_y} ) (\sqrt{u}\sqrt{\omega_y} -1)\right]. 
\ee
Eq.~(\ref{eq:pbg}) gives the complete three dimensional phase boundary. 
For $f_y =0$ ($\omega_y =1$) Eq.~(\ref{eq:pbg}) reduces to the following simple 
expression for the critical force when pulling along the $x$-direction
\be \label{eq:pbx}
\omega_x^c = \frac{u (\sqrt{u}-1)}{\sqrt{u}+1}.
\ee
This coincides with the result obtained by transfer matrix methods
\cite{rosa}. However, that  method is not easily generalized to the case 
where force is applied along the $y$-direction. Eq.~(\ref{eq:pbg}), on
the other hand, is sufficiently general to yield an expression for the 
critical force when pulling along the $y$-direction.  
Setting $f_x=0$ (or $\omega_x=1$) yields the critical force 
when pulling along the $y$-axis:
\be \label{eq:pby}
\omega_y^c = \left[ \frac{1+u^2 + \sqrt{1+2u^2 - 4 u^3 + u^4}}{2 u^{3/2}}
\right]^2
\ee
\begin{figure}
\centerline{
\includegraphics[width=8.0cm]{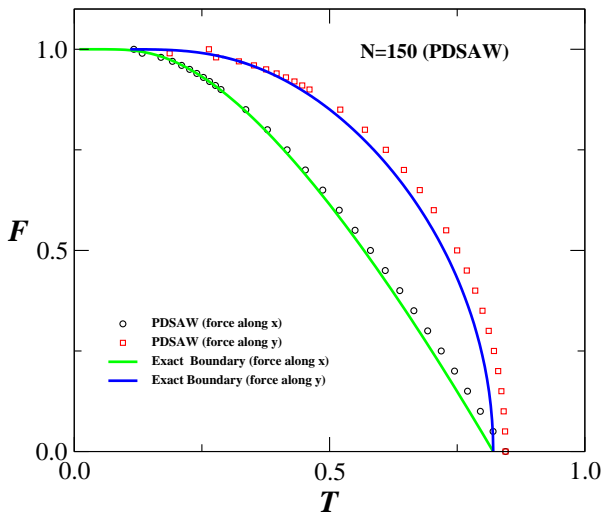} }
\caption{\label{fig:phase} (Color online) The globule-coil phase boundary in the
force-temperature plane.  The response to the force when applied 
along the $y$-direction (squares) is distinctly different 
to the  case where the force is applied along the $x$-direction 
(circles). The continuous 
lines show the exact phase boundaries which are in excellent agreement 
with our numerical results for finite $N = 150$.
}
\end{figure}

For zero force ({\it i.e.} $\omega_x= \omega_y=1$) the expression reduces to
the exact value $ u^c = 3.3829\ldots$ ($T=0.8205 \ldots$ if we set $\eu =1$ ) 
at which thermal unfolding occurs. The exact phase boundaries obtained from 
Eqs.~(\ref{eq:pbx}) and (\ref{eq:pby}) are shown in Fig.~\ref{fig:phase} and
compared to numerical results obtained for $N=150$.
\begin{figure}
\centerline{
\includegraphics[width=2.2in]{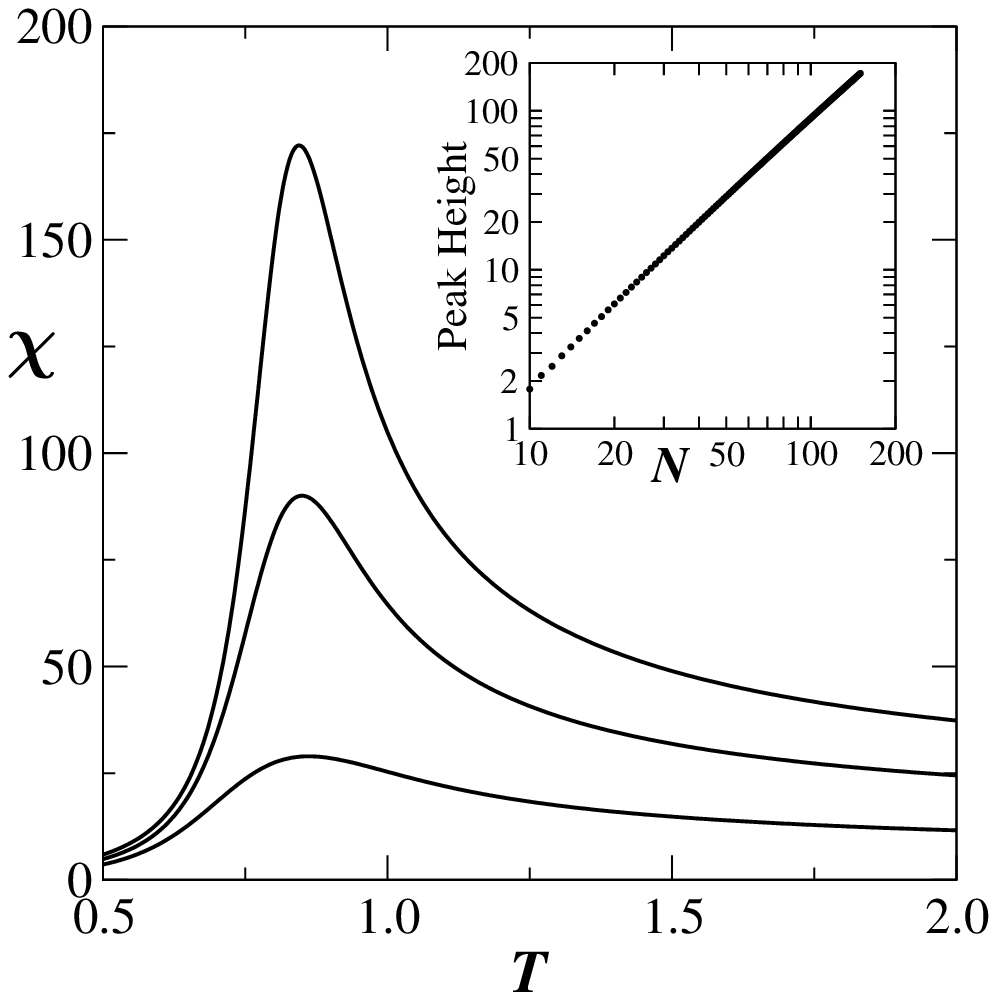}} 
\centerline{\includegraphics[width=2.2in]{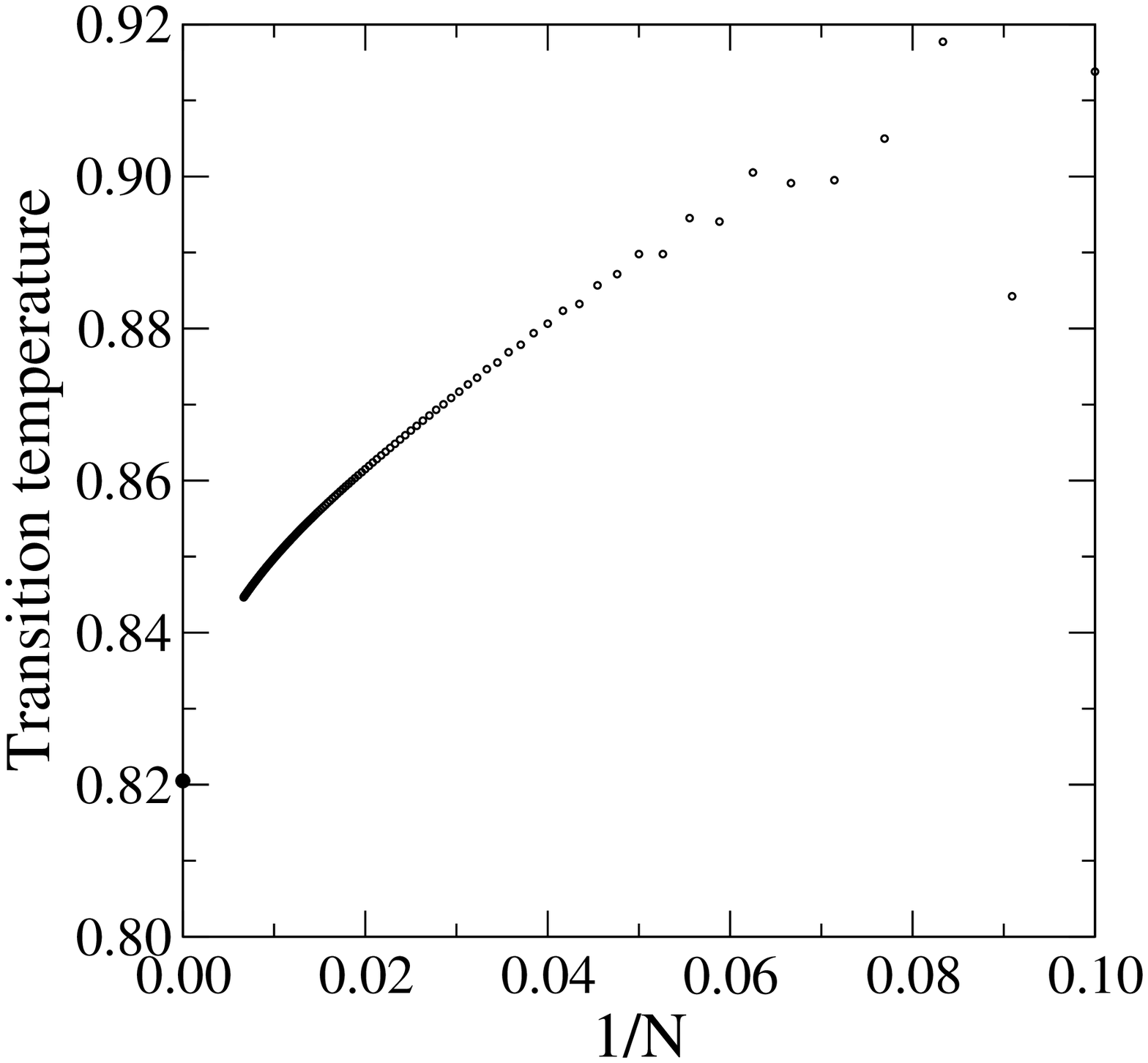} }
\caption{\label{fig:fluc_phaseb} Top: Fluctuations in the number of nearest neighbor 
contacts ($\chi$) as a function of temperature for different $N$. Inset shows how the 
peak height increases with $N$ at the transition temperature; Bottom: The variation of the
transition temperature as $N\rightarrow \infty$. The dot on the $y$-axis marks
the exact transition temperature. }
\end{figure}

\begin{figure*}[t]
\centerline{\includegraphics{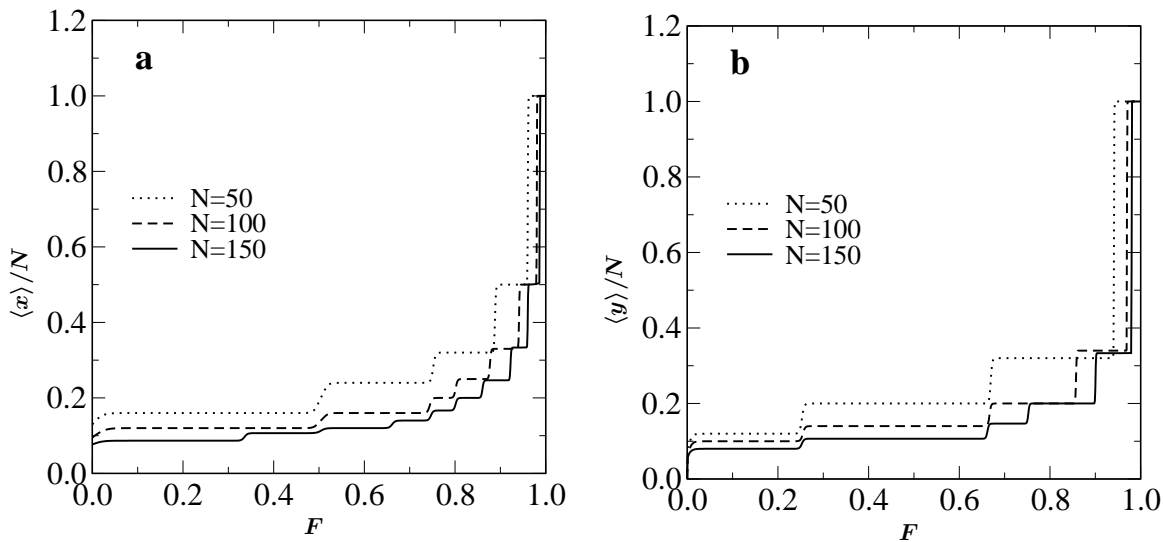}}
\caption{\label{fig:FX_PDSAW} Plots of the average extension $\langle x \rangle/N$ 
($\langle y \rangle/N$) vs the
applied force $F$ for chain lengths $N =150$, 100 and 50 at fixed temperature
$T=0.01$ with the force along the $x$-direction ($y$-direction). 
}
\end{figure*}

\section{Numerical Analysis}

Our numerical studies of the process of force induced protein unfolding  
are based on the exact enumeration of all possible  conformations 
of the chain. This approach not only enables us to obtain precise estimates
for the phase boundary, but also allows us to study the intermediate states
crucial to events occurring during the mechanical stretching process 
\cite{altmann, rief1,schwaiger}.
In this study we have combined the power of 
parallel processing and transfer matrix calculations to extend the
limit of earlier studies not just by one or two monomers but five times over!
Thus we have extended from the previous longest chain length \cite{kumar} $N=30$
to much longer chains of up to $N=150$ steps. 
The partition function is
\begin{equation}
Z_N  = \sum_{N_p, \alpha_x, \alpha_y}
C_N (N_p, \alpha_x, \alpha_y) u^{N_p} \omega_x^{\alpha_x}
\omega_y^{\alpha_y},
\end{equation}
where $C_N$ is the total number of $N$-stepped PDSAWs having $N_p$ non-bonded
nearest neighbor pairs with end-to-end vector  $(\alpha_x, \alpha_y)$.

The value of the transition temperature (for fixed force and finite $N$)  
can be obtained from the fluctuations in the number of non-bonded nearest neighbor 
interactions. The fluctuations are defined as $
\chi=\langle N_p^2 \rangle -\langle N_p \rangle^2$, with the $k$'th moment given by 
\be \label{eq:mom}
\langle N_p^k \rangle = 
\frac{
\sum_{N_p,\alpha_x,\alpha_y} \! N_p^k C_N(N_p,\alpha_x,\alpha_y)  
u^{N_p} \omega_x^{\alpha_x} \omega_y^{\alpha_y}
}{
\sum_{N_p,\alpha_x,\alpha_y} \! C_N(N_p,\alpha_x,\alpha_y)  
u^{N_p} \omega_x^{\alpha_x} \omega_y^{\alpha_y}
}
\ee
When the fluctuations $\chi$ are plotted as a function of temperature
the resulting graph (Fig.~\ref{fig:fluc_phaseb}a) has a peak at the transition temperature. 
The inset shows that the height of the peak in the fluctuation curve grows 
as a power-law with $N$, this being the hall mark of a phase transition.
By setting $\eu =1$ we obtain (for $N=150$)  at zero force ($\omega_x=\omega_y=1$)
the transition temperature $T=0.8446$.
This is already quite close to the value $T=0.8205$ obtained in the
thermodynamic limit ($N\to \infty$). The bottom panel in Fig.~\ref{fig:fluc_phaseb}
shows the variation in the transition temperature plotted against $1/N$ and we see
a pronounced curvature for large $N$ making an extrapolation to the limit $N\to \infty$
very difficult. 

In Fig.~\ref{fig:phase} we show the force-temperature diagrams 
for $N=150$ when pulling along the $x$- and $y$-axis (circles and squares,
respectively).  These numerical estimates for the phase-boundary are in 
excellent agreement with the analytical results (shown as solid lines) 
obtained from Eqs.~(\ref{eq:pbx}) and (\ref{eq:pby}). Thus exact enumeration 
technique works quite well even for small $N$ as far as the phase diagram
is concerned. This has also been seen earlier \cite{mishra} in case of 
the surface adsorption.

\subsection{Constant Force and Constant Distance Ensembles}

Many biological reactions involve large conformational changes  providing 
a well defined mechanical reaction co-ordinate ({\it e.g.} the end-to-end 
distance of a bio-polymer), which has been used in force spectroscopy 
experiments to follow the progress of the reaction \cite{rief,carrion1}. 
Within the constant force ensemble (CFE) the theoretical predictions 
based on exact enumeration results provide qualitative description of the outcomes 
of such experiments \cite{kumar,kumar1} (like the existence of intermediate 
states and different pathways to unfolding). In Fig.~\ref{fig:FX_PDSAW} we have
plotted the average extension as a function of the applied force. 
Fig.~\ref{fig:FX_PDSAW}a shows the result when the force is applied in the
$x$-direction and Fig.~\ref{fig:FX_PDSAW}b when the force is applied in
the $y$-direction. In both cases we see the existence of several  
plateaus showing that there are many intermediate states between the
collapsed and fully extended state. A major difference is that the
number of these intermediate states seems to grow much faster with the
chain length $N$ when pulling in the $x$-direction as compared to pulling
in the $y$-direction. 

\begin{figure}
\centerline{\includegraphics[width=3in]{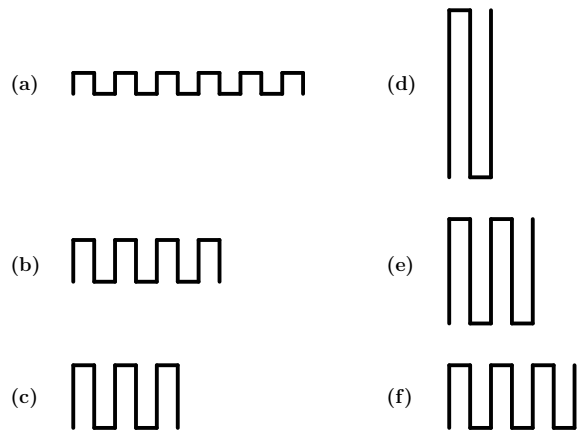}}
\caption{\label{fig:PlatConfig} Schematic diagrams of the configurations 
giving rise to the plateaus in the force-extension curves when pulling in the $x$-direction 
(figures (a)-(c)) and along $y$-direction (figures (d)-(f)).
}
\end{figure}

\begin{figure*}[t]
\centerline{\includegraphics{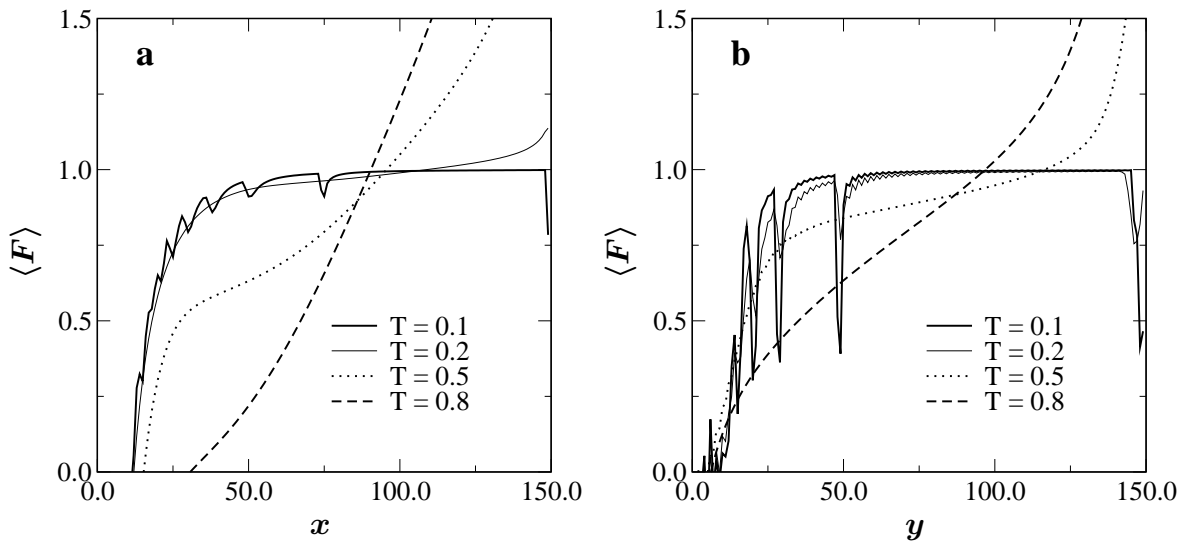}}
\caption{\label{fig:AveF-semi} Plot of the average force 
$\langle F \rangle$ vs the
elongation at different temperatures $T$ for chain length $N =150$. The
marked difference due to a change in pulling direction is clear
from these plots.
}
\end{figure*}

The appearance of the plateaus in the force-extension curves as well as the
differences between the two cases have a simple heuristic explanation. At very
low temperature there are no entropic contributions to the free energy which is
thus dominated by the energetic terms arising from nearest-neighbor interactions 
and the applied force. When pulling in the $x$-direction  we have
to maximize the quantity $G=F x + N_p$, where $N_p$ is the number of contacts.
For $F \to 0$ this means just maximizing $m$, and this is achieved by zigzag 
conformations inscribed in a rectangle which is as close to a square
as possible so that $\langle x \rangle/N$ is close to $\sqrt{N}$. As the
force is increased other configurations may become energetically favored,
essentially corresponding to decreasing the width of the rectangle 
by a unit and increasing the length correspondingly (a more detailed explanation 
can be found in \cite{Guttmann08}; see also \cite{Krawczyk05}). So the plateaus 
seen in Fig.~\ref{fig:FX_PDSAW}a 
arise from inscribing a PDSAW of fixed length $N$ in rectangles, going from right 
to left, of width $w= 2, 3, \ldots \, \sqrt{N}$ as evidenced by the fact that these 
plateaus have height $\langle x \rangle/N=1/w=1/2,1/3,\ldots$. The types of 
conformations giving rise to the right-most plateaus are represented
schematically in Figs.~\ref{fig:PlatConfig}a-c. When pulling in the $y$-direction
we again get a competition between the force which tends to maximize the 
distance of the end-point in the $y$-direction and interactions which try
to inscribe the PDSAW into a square. As illustrated in  Figs.~\ref{fig:PlatConfig}d-f
the conformations giving rise to the right-most plateaus are now those where the
PDSAW is inscribed in a rectangle of length $l=3,5,7,\ldots,  \sqrt{N}$. This
can be seen in  Fig.~\ref{fig:FX_PDSAW}b where these plateaus have height 
$\langle y \rangle/N=1/l= 1/3,1/5,\ldots$. Even lengths cannot occur because 
for these the end-point would have a $y$-component close to 0
giving only a small energetic contribution from the applied force. The upshot
is that for a given length the number of plateaus when pulling in the
$x$-direction will be about twice the number of plateaus  when pulling in the
$y$-direction.

Our longer series data also allows us to go 
a step further in analyzing the model in the constant distance ensemble (CDE).
This is the ensemble best suited to the analysis of experiments performed
using apparatus such as atomic force microscopes. 
In Fig.~\ref{fig:AveF-semi} we have plotted the average force 
$\langle F\rangle$ \cite{kumar1} as a function of the extension when
pulling in either the $x$- or $y$-direction. 
Striking differences between the two cases are obvious from these plots. 
Most obviously we note that at the low temperature $T=0.1$ the force extension 
curve obtained for pulling in the $x$-direction shows unzipping like transition 
characterized by having smooth plateaus. 
However, when the chain is being pulled in the $y$-direction (at the same temperature), 
the force extension curve exhibits ``saw-tooth'' like oscillations indicating that the 
transition is akin to shearing.
This clearly demonstrates that even the nature of the unfolding transition 
can change dramatically solely by varying the pulling direction. 
\begin{figure}
\centerline{\includegraphics[width=2.4in]{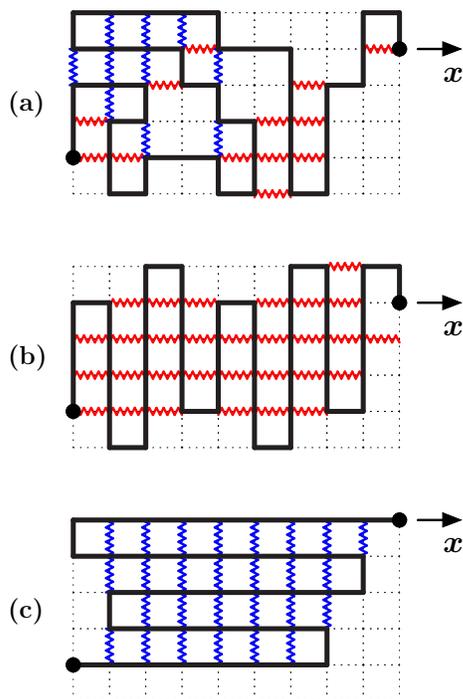}}
\caption{\label{fig:asasaw} (Color online) Schematic representation of ASASAWs on the
square lattice. (a) Different nearest neighbor interactions along the
$x$- and $y$-directions introduce anisotropy in the model polymer. (b)
represents ground state conformations which dominate the partition function
similar to Fig.~\ref{fig:pdsaw} when subjected to force along the $x$-direction. 
(c) Same as (b) but resembles Fig.~\ref{fig:pdsaw}  when subjected to force along the 
$y$-direction.
}
\end{figure}

\section{Anisotropic-Self-Attracting-Self-Avoiding Walks}

The model described above has an exact solution and gives qualitative
features similar to experiments. However, the physical
constraints imposed by experimental set-ups has not been taken fully
into account. For example, in experiments using atomic force microscopes, 
receptor and ligand molecules are attached to a substrate and a transducer,
respectively, which introduce anisotropy in the systems. 
Also steric constraints due to the confinement imposed by the
experimental set-up  can lead to a loss of entropy and thus result in 
modifications to the behavior of the chain.
Furthermore the constraint of partially directing the walk so it cannot take steps
in the negative $x$-direction is not really appropriate to most experimental
conditions. Such a constraint is typically only appropriate when the system is under
flow or constant external field \cite{lemak,lemak1}.  A  more realistic model of
polymers  is provided by self-attracting-self avoiding walks (SASAWs),
which is well suited to the modeling  of a linear chain 
in a poor solvent \cite{degennes, vander}. However, this model is isotropic 
in nature. In this paper we introduce 
anisotropy to the model.  We achieve this  by considering nearest neighbor
interactions with different strengths along the $x$- and $y$-directions as 
shown in Fig.~\ref{fig:asasaw}a. This is in accordance with
real proteins where the interactions along different directions can differ 
by orders
of magnitude \cite{creighton}. Hence by changing the strength of the interactions 
one can vary the degree of anisotropy in the system and study the effect
this has on the force-temperature phase diagram. The partition function for 
anisotropic-self-attracting-self-avoiding walks (ASASAWs) may be written as  
\begin{equation}
  Z_N^{\prime}  = \sum_{(N_{px}, N_{py},
    | \alpha |)} C_N (N_{px}, N_{py}, | \alpha |) {u_{x}}^{N_{px}}
   {u_{y}}^{N_{py}} \omega_\alpha^{| \alpha |}.
\end{equation}
Here $C_N$ is the total number of ASASAWs
of $N$ steps having $N_{px}$ and $N_{py}$ nearest neighbor
pairs along the $x$- and $y$-directions, respectively, while 
$u_x =\exp[-\beta \epsilon_x]$ ($u_y =\exp[-\beta\epsilon_y]$) are 
the Boltzmann weights associated with the
nearest neighbor interactions between non-bonded monomers
along, respectively, the $x$- and $y$-directions. 
\begin{figure}
\centerline{\includegraphics[width=3in]{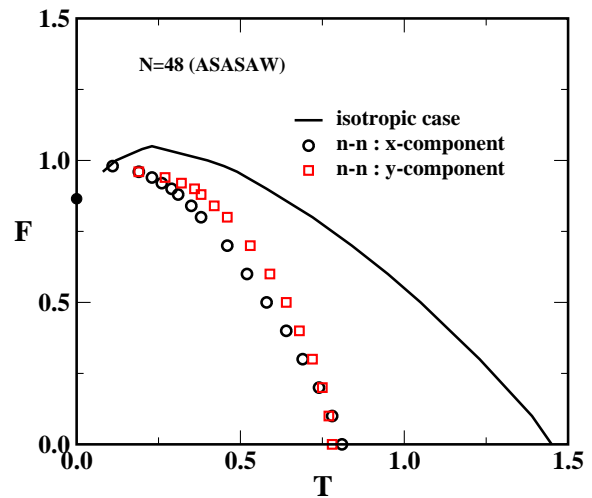} }
\caption{\label{fig:PDflex} (Color online) 
The force-temperature phase diagram for ASASAWs. The solid line
represents the phase boundary for the isotropic SASAWs.
}
\end{figure}

\begin{figure*}[t]
\vspace{5mm}
\centerline{\includegraphics{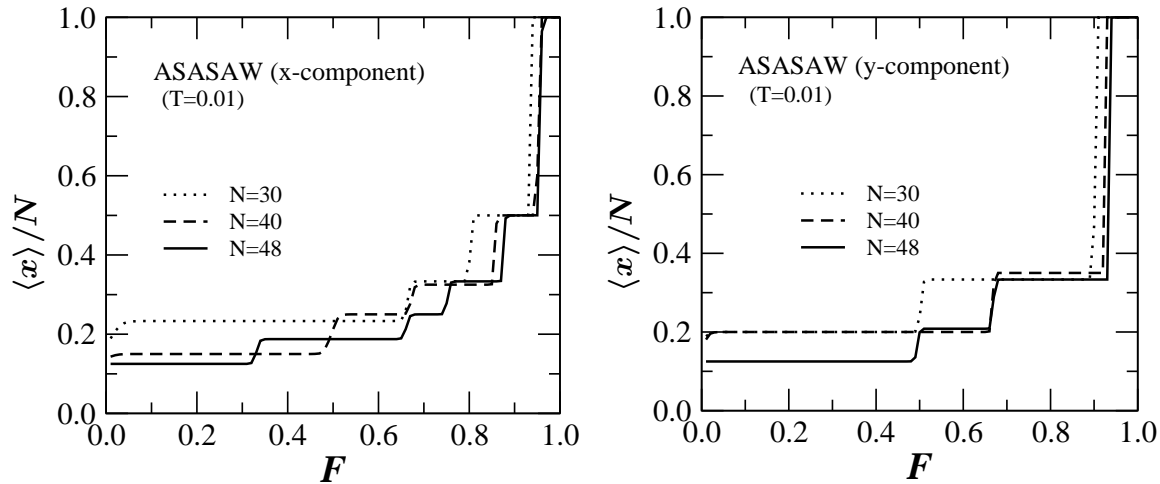}} 
\caption{ \label{fig:AveX}
Left panel:  The average scaled elongation $\langle x \rangle/N$ as a function of the
force with interactions along the $x$-direction at fixed low temperature $T=0.01$ for 
different chain lengths;  Right panel: Same as the left panel but for interactions 
along the $y$-direction.
}
\end{figure*}

\begin{figure}
\includegraphics[width=3.5in]{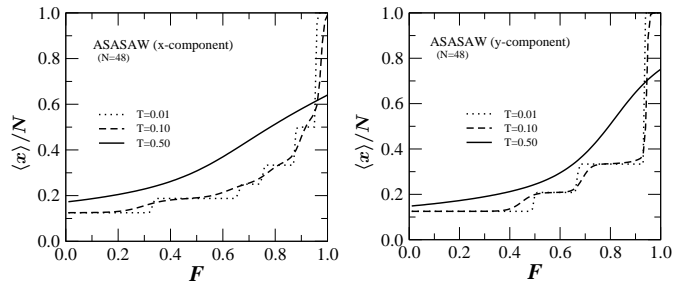} 
\caption{\label{fig:AveXvsT} Left panel: The average scaled extension $\langle x \rangle/N$ 
along the $x$-direction as a function of the force at different temperatures with
chain length $N = 48$; Right panel: Same as the left panel but with interactions 
along the $y$-direction.}
\end{figure}

\begin{figure}
\includegraphics[width=3.5in]{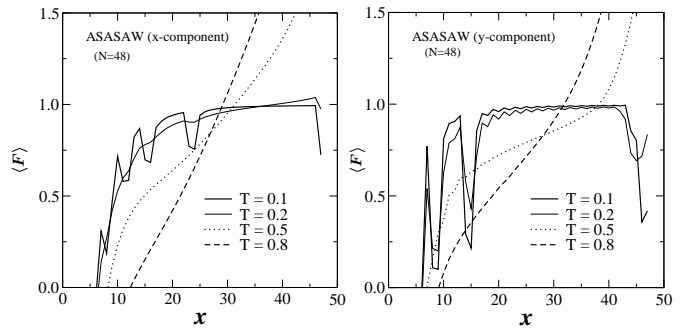} 
\caption{\label{fig:AveFvsT}  Left panel: The average force $\langle F \rangle$ 
vs the extension $x$ with interactions along the $x$-direction alone at different 
temperatures with chain length $N = 48$;  Right panel: Same as the left panel but 
with interactions along the $y$-direction.}
\end{figure}

The qualitative as well as quantitative features of the force-temperature phase 
diagram as shown in Fig.~\ref{fig:PDflex} for the isotropic case ({\it i.e.}  
$u_x = u_y =u$)  remain the same as those reported in \cite{kumar1}.  
To make our model closely resemble the PDSAW discussed above, we set either $u_x = 1$ or 
$u_y = 1$ and apply a force along the $x$-direction. For $u_y = 1$ the ground state
is dominated by configurations similar to the one shown in Fig.~\ref{fig:asasaw}(b).
This closely resembles the PDSAW model (Fig.~\ref{fig:pdsaw}) subjected to a force 
along the $x$-axis and the nature of the transition is akin to an unzipping 
transition. On the other hand for $u_x =1 $ the ground state is dominated by 
configurations like the one shown in Fig.~\ref{fig:asasaw}(c). This closely 
resembles the configurations of Fig.~\ref{fig:pdsaw} with a force along the $y$-axis 
and consequently the nature of the 
transition is akin to a shearing type transition. It should be noted that the number
of ASASAWs, $C_N$, is different from the number of PDSAWs even if we put $u_y = 1$. 
We observe a significant decrease in the transition temperature compared
to SASAWs. This is caused by  a decrease in entropy which allows 
the polymer to acquire conformations similar to a $\beta$-sheet.
The effect of surface confinement is also evident from these plots.
The transition temperature is found to be less for $u_x = 1$ in comparison
to $u_y =1$. Another major difference between the ASASAW and SASAW phase-diagrams 
is that there appears to be no re-entrance in the ASASAW model when $u_x =1 $ or $u_y =1$.
It is well known that the re-entrant behavior of the SASAW model is due to a non-zero
ground state entropy, that is, the number of configurations in the ground state grow
exponentially with $N$ (these configurations are just Hamiltonian walks inscribed in
a rectangle with side lengths as close as possible to $\sqrt{N}$). For the ASASAWs we
study here the ground-states are, as argued above, folded sheets
as in Fig.\ref{fig:asasaw}(b) and (c), respectively. The number of such configurations 
do not grow exponentially with $N$ and are thus too few to give a non-zero ground state entropy,
which is why we do not see re-entrance. However, if both $u_x$ and $u_y$ 
are greater than 1  then one may see a re-entrant phase diagram.

The force extension curve shown in Fig.~\ref{fig:AveX} not only shows the existence 
of intermediate states but also the emergence of new states depending on
the direction of the interactions in accordance with the findings for PDSAWs. 
As the temperature is increased we observe in Fig. 10 (as expected) that the intermediate 
states are washed out due to the resulting increase in entropy. 
We also observe that the system attains a higher stability 
against the force when the interactions are along the $y$-direction 
and that intermediate states survive
even at high temperatures as may be seen from Fig.~\ref{fig:AveXvsT}.

Finally we analyze the ASASAW model in the constant distance ensemble.
In Fig.~\ref{fig:AveFvsT} we have plotted the average force 
$\langle F\rangle$ \cite{kumar1} as a function of the extension 
when pulling in the $x$-direction with interactions either along 
the $x$- or $y$-direction.  The results are qualitatively very similar to 
those obtained for the PDSAW model (see Fig.~\ref{fig:AveF-semi}). In 
particular we note that the oscillation at low-temperature are much more 
pronounced in the case where the interactions are in the $y$-direction.
As explained above (see Fig.~\ref{fig:asasaw}) there is a close correspondence
between the ground-state (low-temperature) configurations of the ASASAW 
with interactions in the $x$- or $y$-directions and the PDSAW with a force 
applied along the  $x$- or $y$-direction, respectively. This 
close correspondence is reflected in the similarity of the results.

\section{Conclusion}

In summary we have demonstrated via the analytical solution 
for the phase boundary of PDSAWs as well as through 
high precision numerical calculations that changing the 
pulling direction can change the nature of the unfolding transitions and 
give rise to very different phase diagrams. The emergence of many new
intermediate states suggests that there may be many pathways to the
unfolding of a polymer. The findings for the PDSAW model 
have been confirmed by our study of ASASAWs. The inclusion of anisotropy 
in the traditional SASAW model of polymers gives unequivocal
evidence that the features observed here for the PDSAW model are in
fact  generally true and not an artifact of the PDSAW model. In future additional 
work like extensive Monte Carlo simulations and molecular dynamics studies are
needed to better understand the role played by anisotropy and how varying the 
pulling direction can help us gain a better and deeper understanding of biomolecules.

\section*{Acknowledgments}

This research has been supported by the Department of Science and Technology 
and University Grants Commissions, India and also by the Australian Research
Council.  The calculations presented in this 
paper used  the computational resources of the Australian Partnership for 
Advanced Computing (APAC) and the Victorian Partnership for Advanced 
Computing (VPAC).


\begin{thebibliography}{99}
\bibitem{suppl}The full details of the calculations will be given elsewhere.
\end{thebibliography}
\end{document}